# Benchmarking network fabrics for data distributed training of deep neural networks

Siddharth Samsi, Andrew Prout, Michael Jones, Andrew Kirby,
Bill Arcand, Bill Bergeron, David Bestor, Chansup Byun, Vijay Gadepally,
Michael Houle, Matthew Hubbell, Anna Klein, Peter Michaleas, Lauren Milechin,
Julie Mullen, Antonio Rosa, Charles Yee, Albert Reuther, Jeremy Kepner

MIT Lincoln Laboratory, Lexington, MA

*Abstract*—Artificial Intelligence/Machine Learning applications require the training of complex models on large amounts of labelled data. The large computational requirements for training deep models have necessitated the development of new methods for faster training. One such approach is the data parallel approach, where the training data is distributed across multiple compute nodes. This approach is simple to implement and supported by most of the commonly used machine learning frameworks. The data parallel approach leverages MPI for communicating gradients across all nodes. In this paper, we examine the effects of using different physical hardware interconnects and network-related software primitives for enabling data distributed deep learning. We compare the effect of using GPUDirect and NCCL on Ethernet and OmniPath fabrics. Our results show that using Ethernet-based networking in shared HPC systems does not have a significant effect on the training times for commonly used deep neural network architectures or traditional HPC applications such as Computational Fluid Dynamics.

## I. INTRODUCTION

High Performance Computing (HPC) centers have traditionally focused on large scale applications developed in high performance languages such as C and Fortran that leverage MPI for scaling up. Over the past several years, these workflows have evolved to become significantly more diverse and have required HPC systems to evolve along with them. High level programming languages such as MATLAB, Python and Julia are being increasingly used for a variety of scientific applications including Artificial intelligence (AI). As AI has become an increasingly larger component of the HPC workload, enabling the continuum of performance and high productivity by choosing the right combination of performance and productivity tools has become more complex and critical at the same time.

Deep Neural Networks (DNN) have been applied successfully in many diverse domains such as image classification, video analysis, language modeling and translation, medical imaging and weather [1]–[4]. The effective training of DNNs requires large amounts of data. As models have gotten deeper and wider, the number of unknown parameters for a network have increased exponentially. For example, the VGG16 [5] model has almost 140 million trainable parameters. The combination of large amounts of data and increasingly large model sizes leads to very long training times for these models. Table I shows the time required to train commonly used networks. The training of deep neural networks is computationally expensive primarily due to two factors: the number of layers and the configuration of each layer, which results in a large increase in the number of add/multiply operations in the forward and backward pass. The second major cause of large computational costs is the size of the training dataset. This does not take into account the number of hyperparameter searches that are required to identify the model configuration required to achieve acceptable performance on a particular task. Thus, significant amount of research has focused on accelerating the training of deep neural networks.

Table I: Training time for deep neural networks.

| Model Name | Training Time | Hardware Used |
|---|---|---|
| AlexNet [6] | 5-7 days | 2 x NVIDIA GTX 580 |
| InceptionV3 [7] | 2 weeks | 8 x NVIDIA Tesla K40 |
| ResNet50 [8] | 29 Hours | 8 x NVIDIA Tesla P100 |
| VGG16 [5] | 2-3 weeks | 4 x NVIDIA Titan Black |

Strategies for accelerating the training of DNNs span across a wide range from custom hardware to scaling up using High-Performance Computing (HPC) resources. While a wide variety of custom hardware is available specifically for DNN training and inference [9], training DNNs is becoming an increasingly larger part of HPC workloads. There have been many published works that discuss the distributed training of deep neural networks. Goyal e.t al. [10] used 256 GPUs to train ResNet-50 on the ImageNet dataset in 1 hour. Yang et.al. [11] trained AlexNet on the same dataset in 11 minutes on 2,048 Intel Xeon Platinum 8160 processors and ResNet-50 in 20 minutes using 2,048 Intel Xeon Phi 7250 processors. Both of these implementations used the data distributed approach to parallel training. Yin et. al. [12] details scaling strategies and optimizations for training deep neural networks on 6,144 NVIDIA V100 GPUs. Torsten et. al. [13] provides a comprehensive review of DNN architectures and parallel training

This material is based upon work supported by the Assistant Secretary of Defense for Research and Engineering under Air Force Contract No. (FA8721-05-C-0002 and/or FA8702-15-D-0001). Any opinions, findings and conclusions or recommendations expressed in this material are those of the author(s) and do not necessarily reflect the views of the Assistant Secretary of Defense for Research and Engineering.

strategies.

The rest of the paper is organized as follows. Section II discusses the HPC system on which these tests were run. Section III describes the benchmarks used to test the effect of communication networks on application performance, current approaches to data parallel distributed learning and published results. We discuss our results in Section IV and Section V offers some conclusions.

## II. SYSTEM ARCHITECTURE

For over a decade, the MIT Lincoln Laboratory Supercomputer Center (LLSC) has enabled interactive, on-demand high performance computing [14], [15] by integrating high-productivity tools with traditional HPC systems to enable the prototyping and development of a wide range of scientific applications. As machine learning (ML) workflows have occupied an increasingly large proportion of the types of jobs running on the system, we have developed new tools that support interactive web-based environments [16] such as Jupyter notebooks for prototyping machine learning applications. In addition to the changes in system software required to support these use cases, our hardware and systems architecture considerations must include the simultaneous support of general-purpose HPC workloads and the optimized, often accelerator-driven compute resources required by AI/ML applications. From the perspective of computations, GPUs are predominantly used for AI/ML applications. Coupled with high-end processors, such a node configuration can well serve both requirements. Another critical component of an HPC system is the networking hardware and can potentially have a significant impact on distributed application performance. The focus of this paper is on the effect of two commonly deployed networking technologies - Ethernet and Omnipath.

### A. TX-GAIA (Green AI Accelerator)

The experiments described in this paper were conducted primarily on the MIT Lincoln Laboratory Supercomputing Center's TX-GAIA system. A notional depiction of the system's layout is shown in Figure 1. The TX-GAIA HPC system consists of 448 compute nodes, each with two 2.5 GHz Xeon Gold 6248 processors and 384GB of RAM. All nodes have two NVIDIA Tesla V100 GPUs with 32GB of GDDR5 RAM each. Both GPUs are connected via PCIe slots to CPU1 as shown in the node-level topology in Figure 2. Each node on the system has two independent back-end fabrics: a 100 Gb/s Intel Omnipath as well as a 25 Gb/s Ethernet interconnect using Mellanox ConnectX-4 adapters with all servers connected to a single, non-blocking Arista DCS-7516 Ethernet core switch. The GPUs, Omnipath, and Ethernet cards were all connected to PCIe slots routed directly to the Xeon processors without any intermediary PCIe switches.

### B. RDMA, RoCE and GDRDMA

The Remote Direct Memory Access (RDMA) protocol is the backbone of parallel applications requiring low-latency, high-bandwidth communication of data between clustered nodes in a high performance computing system, providing a low-overhead, direct zero-copy mechanism for accessing remote main system memory (RAM) without the involvement of the remote host's CPU or operating system kernel.

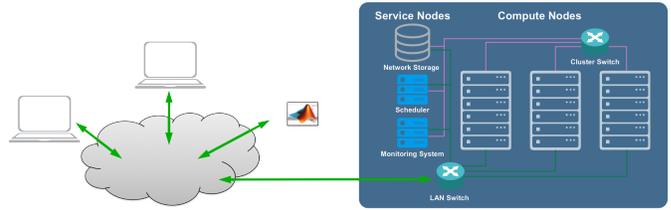

Fig. 1: Architecture of the MIT SuperCloud systems. Users connect to the system over either a local area network or a wide area network. At the time of connection, their system joins the MIT SuperCloud and can act as a compute node in order to run parallel programs interactively. The centerpiece of the MIT SuperCloud is several file systems (Seagate, DDN, Dell, Hadoop, and Amazon S3) running on several different network fabrics (10/25 GigE, InfiniBand, OmniPath). The MIT SuperCloud is representative of a supercomputer designed for interactivity and rapid prototyping.

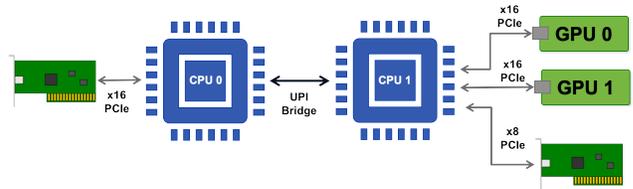

Fig. 2: PCI topology of individual nodes in the Tx-GAIA system. UPI: Intel Ultra Path Interconnect.

Traditionally, high-performance computing systems have relied on variants of the InfiniBand protocol to provide the fabric over which these RDMA implementations communicate; however, continued improvements in the stability, reliability and latency of commodity Ethernet networking hardware have significantly narrowed the best-case performance gap between a traditional InfiniBand fabric and one backed by low-latency non-blocking Ethernet switches. The RDMA over Converged Ethernet (RoCE) protocol was conceived to take advantage of these improvements, allowing supercomputers and datacenters to simplify their network topologies by unifying networking, compute and storage atop a single, shared fabric.

The rise of offload accelerators such as GPUs and FPGAs used in highly parallel workloads has necessitated the development of technology to allow RDMA to facilitate data movement to and from locations other than main system memory, allowing similar direct communication between the memory spaces of hardware accelerators situated in physically separate hosts. The GPUDirect technology developed by NVIDIA is one such implementation. GPUDirect Peer-to-Peer provides a pathway for direct communication between peer Kepler-class (K-Series) or newer GPUs in the same system over the PCI Express bus, completely offloading the CPU and host operating system kernel. GPUDirect RDMA (GDRDMA) provides the same CPU bypass for RDMA requests accessing accelerators in physically separate hosts over a network interconnect or fabric. The benefits of GPUDirect are dependant on the exact system configuration, such as affinity of the PCIe slots when more than one CPU socket is present and whether the system architecture includes one or more PCIe switches, requiring additional planning in the system design for optimal performance.

This burst of recent developments in computer networking communications protocols and technologies compels robust benchmarking efforts to evaluate their potential. Our specific interests were to assess the impact of network protocol and back-end network fabric selection on a range of common HPC problems.

*C. Scaling Deep Learning on HPC systems*

Training deep learning (DL) models requires significant compute time as shown in Table I. Common approaches to distributed training of deep neural networks includes data parallel, model parallel and pipeline parallel implementation of training procedures. For a detailed review of these approaches, readers are referred to Ben-Nun and Hoefler [13]. Yin, et. al [12] describe best practices for scaling deep learning at extreme scale on the leadership-class Summit supercomputer. The experiments described here adapted the recommended optimizations in [12]. The benchmarks used in this paper were implemented in Tensorflow and are publicly available from the Tensorflow git repository here https://github.com/tensorflow. The version of Tensorflow used in these experiments was 1.14. The Horovod [17] framework was used to enable data distributed training. Horovod was configured to use MPI, NCCL and GPUDirect over OmniPath and Ethernet.

### III. BENCHMARKS DESCRIPTION

We characterized the performance of the system using two broadly different benchmarks. The first set of benchmarks used were the Tensorflow benchmarks (available [18]), and the second was a CFD application written using C and MPI.

*A. Deep Learning Benchmarks*

There are three commonly used approaches for distributed training of deep neural networks:

1) Data Parallel: In this approach, the model is replicated across multiple compute nodes and the training data is distributed across all the node. Each processor trains on a subset of the total dataset, which is the primary source of performance gains when training in this manner. At the end of each epoch, all parallel processes must communicate with each other to synchronize gradients. Thus, as we scale up to larger numbers of processors, the amount of compute per epoch decreases while the communication time increases, leading to sub-optimal speedups. However, this is an effective strategy for training models on large amounts of data and has been shown to scale up to thousands of CPUs [11] and GPUs [12].

2) Model Parallel: In the model parallel approach, the model to be trained is split across multiple processors and the same training data is used on all processors. This is typically beneficial when the model is very large but has the drawback of increased communication between the processors. Since each layer in a neural network depends on the output of the previous layer, communication costs can be significant. This strategy is not as commonly used as data parallelism.

3) Pipeline Parallel: This strategy aims to avoid communication overheads from data parallelism [19] by partitioning model layers across processors. By using multiple mini-batches at a time, the compute units are kept busy and communication can be amortized over the computation time. Harlap et. al. show that this strategy can reduce overall communication costs at the expense of an increased memory footprint. For details, we refer the reader to [19] and [13].

In this paper, we use the data parallel strategy for benchmarking purposes. We test data distributed training for both TensorFlow and PyTorch on two different network interconnects. The open-source TensorFlow benchmarks available from [18] provide implementations of ResNet50, ResNet50_v1.5, VGG16, InceptionV3 and several others and were used for testing. These implementations provide a simple way to test the software and hardware infrastructure for deep learning deployment. Tuning knobs include the ability to use full or mixed-precision training, Horovod or native TensorFlow based distributed training, configurable all-reduce and all-gather mechanisms, among many other options. In this paper, our focus is on comparing the effect of using 25-Gigabit Ethernet and 100-Gigabit Omnipath networks on training times. As a result, while the actual implementation of these benchmarks may have opportunities for improvement, they provide a baseline for comparing different networking hardware and their effect on training time.

*B. Computational Fluid Dynamics*

The Computational Fluid Dynamics solver is known as CartDG [20]–[22] which is a Discontinuous Galerkin method that discretizes the compressible Navier-Stokes equations. CartDG has been used in the high-fidelity simulation of wind energy applications [23], [24] and in aerospace applications including rotorcraft wakes [25]. The solver is designed for computational efficiency on Cartesian meshes and uses tensor-product, collocation-based basis functions to achieve over 10% peak of theoretical compute performance [26]. The software implements the Message Passing Interface (MPI) for distributed-memory computation by equally partitioning the Cartesian mesh into identical mesh blocks and enables computation-communication overlap to hide communication latency. It has been demonstrated to scale to over one million MPI ranks [27] on the ALCF Mira Supercomputer.

## IV. RESULTS

All benchmarks were run on the same compute infrastructure with 25-Gigabit Ethernet and 100-Gigabit Omnipath network interfaces. Only one interface was used at a time.

### A. CartDG

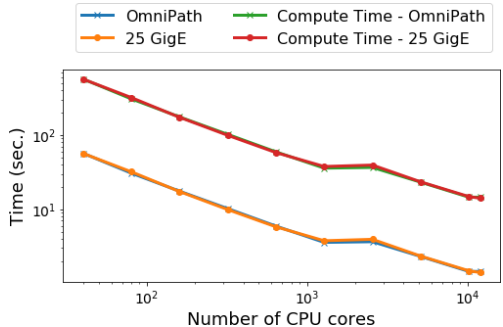

Fig. 3: Communication and compute time for the CartDG CFD application: The application was run on CPU cores and did not use GPUs for compute. In the case of both 25-Gigabit Ethernet and 100-Gigabit Omnipath, the communication time is identical. Using an Ethernet-based communication fabric on a shared HPC system does not adversely affect the scaling of this traditional HPC application.

To test the various network communications, the CFD solver CartDG is tested using a parallel strong-scaling benchmark which partitions a computational mesh of fixed size over varying CPU core counts. The distributed-memory communication is set up such that each CPU core contains identical computational work and communication patterns. This is done by equal mesh partitioning and simulation of a three-dimensional periodic Cartesian domain. The problem chosen for this test is composed of 83,886,080 unknowns on a 32x32x32 mesh.

Figure 3 shows the results of benchmarking the CartDG application on Ethernet and Omnipath networks. As seen in the Figure, the software parallel strong scales well as more CPU cores are introduced indicating the simulation time is dominated by the computational work compared to the communication. Notice that the communication times for the OmniPath and 25 GigE interconnects are nearly identical reflecting the minimal impact of the network. An interesting artifact appears in the strong scaling test for both the OmniPath and 25 GigE interconnects: the compute and communication times plateau from 1,280 cores to 2,560 cores, then continues decreasing linearly on a secondary trend to over 12,000 cores. This is due to the node placement within a single rack; there are 32 nodes in a rack which corresponds to 1,280 cores. Thus, communication performance is impacted for the MPI packets between system racks.

### B. Distributed Deep Learning

We benchmark ResNet50, ResNet50_v1.5, InceptionV3 and VGG16 networks on both network fabrics. In our testing, only one network fabric was enabled at a time when running these benchmarks. All models were trained on the ImageNet [28] dataset. Training data was first converted to the TFRecord format and stored in 1024 individual files. The use of larger files for the training data also ensured that we minimized any slowdowns due to large scale reads from the shared central Lustre file system. TensorFlow implementations of these models used Horovod [17] compiled with the NVIDIA NCCL library was used for distributed training. We used CUDA-aware OpenMPI [29] 4.0.1 for parallel process launches. In the case of the 25Gigabit Ethernet tests, the RDMA-over-Ethernet protocol was enabled for communications by setting appropriate flags during MPI launch. Figure 4 shows the results of running these benchmarks over the two network fabrics. Across all tests we found that the Ethernet-based fabric suffered an average reduction of 12.78% images per second as compared with the Omnipath network. In addition, we also test the effect of both network fabrics on different all-reduce strategies. Figure 5 shows the results of this comparison. In all cases, we see consistent performance across GigE and Omnipath networks.

We also evaluated the effect on processor affinity of the network cards. Small scale tests were done with the following configurations of PCIe lane affinity:
1) Both GPU cards and the Mellanox Ethernet card to **CPU1**, with Omnipath card to **CPU0**.
2) One GPU card each to **CPU0** and **CPU1**.
3) Both GPU cards and the Omnipath card to **CPU1**, with Mellanox Ethernet card to **CPU0**.

No statistically significant difference could be detected between these configurations, and in the end TX-GAIA was built with the first configuration listed above.

## V. CONCLUSION

In this paper we describe experiments comparing 25 Gigabit Ethernet and 100 Gigabit Omnipath based network topologies for building a shared High Performance Computing system. We evaluate commonly used deep learning benchmarks implemented in TensorFlow and PyTorch and show that, for these workloads, the use of an Ethernet-based interconnect leveraging RDMA over Converged Ethernet does not adversely affect the overall application performance when scaling up to 512 GPUs. Similarly, we also evaluate a more traditional HPC workflow in the form of a CFD application and show that its overall performance remains roughly identical regardless of back-end fabric selection.

It is undeniable from existing literature that specialized high-speed interconnects can provide better performance, touting lower latency and higher effective bandwidth than Ethernet. However, the gap between the real-world application performance of these specialized interconnects and generally available Ethernet is not so large as to be fatal to its use in high performance computing and moderate-scale machine learning applications. Overall system cost, ease of management/maintenance and availability of trained personnel are also significant factors in technology selection. The traditional TCP/IP/Ethernet framework is almost universally adopted in modern computer systems, has impressive robustness and a rich ecosystem of widely-available and widely-understood monitoring, debugging and troubleshooting tools. The prevalent use of Ethernet in enterprise networking ensures vendor support for remote operation, while, in our experience, newer and more specialized fabric requires frequent, hands-on physical intervention to resolve issues, introducing considerable administrative overhead and delay for a lights-out, remotely-operated datacenter such as the one housing TX-GAIA. Our results have shown that performance can be achieved with a well-designed Ethernet fabric that nearly matches the performance of more specialized fabrics for many workloads. The ability to reinvest the savings realized in hardware and personnel, combined with the increased reliability and availability of the system, can be a worthy trade-off for many organizations that understand their workload.

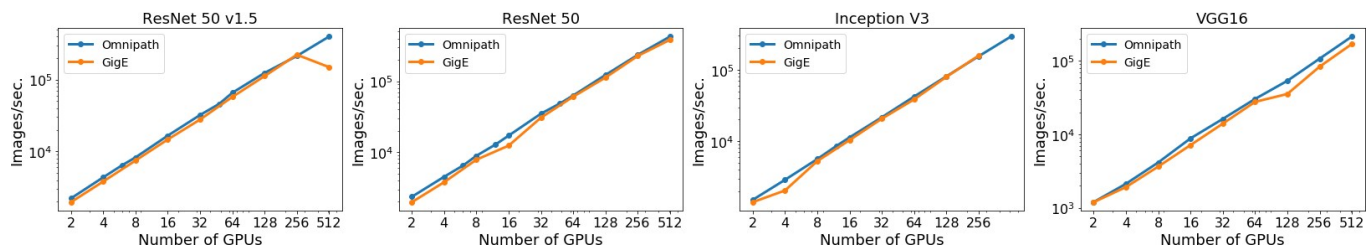

Fig. 4: Comparison of 25GigE and 100Gb Omnipath networks for distributed training of convolutional neural networks: Figure show the images per second processed for ResNet50 (v1.5), VGG16 and the InceptionV3 networks. We see a small degradation in performance with 25Gig Ethernet based network as compared with a 100Gb Omnipath.

ACKNOWLEDGMENTS

The authors acknowledge the MIT Lincoln Laboratory Supercomputing Center for providing HPC resources that have contributed to the research results reported in this paper. The authors wish to acknowledge the following individuals for their contributions and support: Bob Bond, Steve Rejto and Dave Martinez.

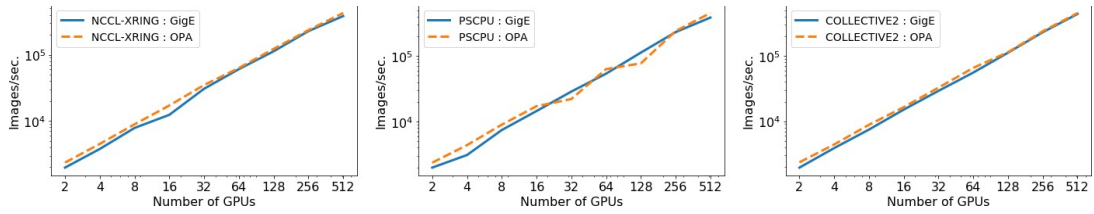

(a) ResNet50: In the case of both network fabrics, performance is seen to increase linearly for the all-reduce strategies considered here.

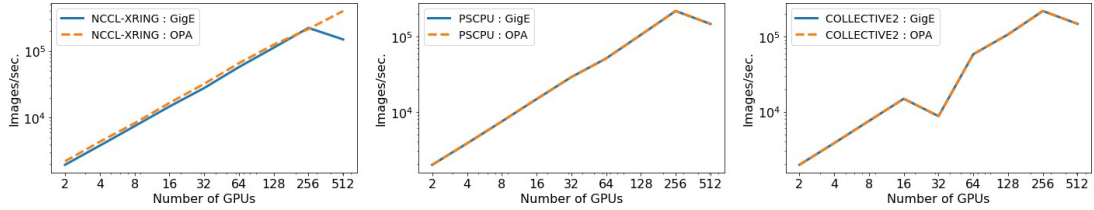

(b) ResNet50 v1.5: Both networks show comparable performance for each all-reduce strategy. In the case of the COLLECTIVE2 strategy, a degradation in performance is see for both network fabrics when training on 32 GPUs. While this needs additional investigation to determine the cause, simply switching to a different all-reduce algorithm avoids this issue.

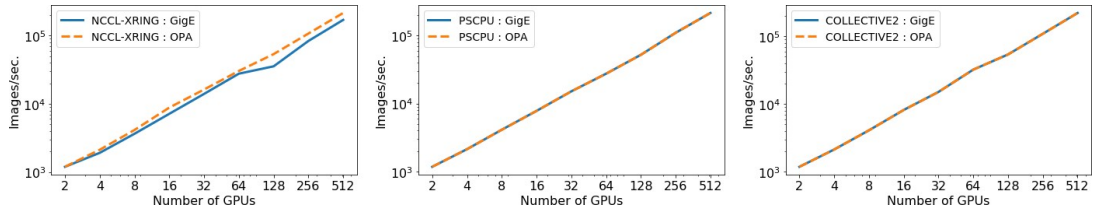

(c) VGG16

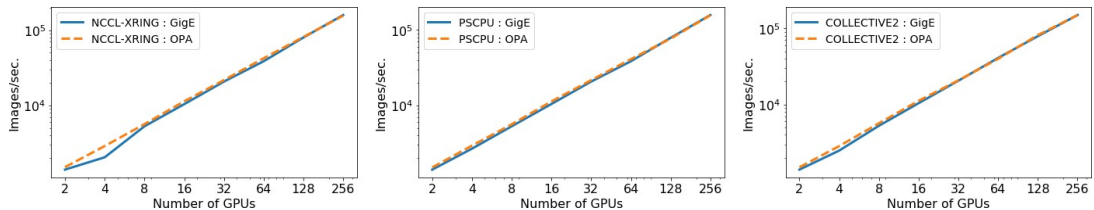

(d) Inception V3

Fig. 5: Comparison of all-reduce algorithms for distributed training of ResNet50, ResNet50 v1.5, VGG16 and Inception V3: Three different all-reduce strategies were compared on GigE and Omnipath network fabrics. All models were trained on the same hardware platform using the same software stack, including benchmark implementations. In all cases, the performance of both network fabrics is observed to be similar at least through 256 GPUs. The reason for the performance degradation for the ResNet50 v1.5 at 512 GPUs needs further investigation, but is likely due to bandwidth requirements growing at the increases scale, saturating the 25Gbit/sec of bandwidth available on our Ethernet interconnect.